
\input harvmac
\Title{\vbox{\baselineskip12pt\hbox{CERN-TH.6906/93}\hbox{DFUPG 80/93}}}
{\vbox{\centerline{Topological Excitations in}
\vskip2pt\centerline{Compact Maxwell-Chern-Simons Theory}}}
\centerline{M. C. Diamantini and P. Sodano}
\medskip
\centerline{Dipartimento di Fisica e Sezione I.N.F.N., Universit\'a di
Perugia}\centerline{06100 Perugia, Italy}
\bigskip
\centerline{C. A. Trugenberger}
\medskip
\centerline{Theory Division, CERN}
\centerline{1211 Geneva 23, Switzerland}

\vskip .3in
We construct a lattice model of compact (2+1)-dimensional
Maxwell-Chern-Simons theory, starting from its formulation in terms of
gauge invariant quantities proposed by Deser and Jackiw. We thereby
identify the topological excitations and their interactions.
 These consist of monopole-antimonopole pairs
bounded by strings carrying both magnetic flux and electric charge. The
electric charge renders the Dirac strings observable and endows them with
a finite energy per unit length, which results in a
linearly confining string tension.
Additionally, the strings interact via an imaginary, topological
term measuring the (self-) linking number of closed strings.

\Date{CERN-TH.6906/93, DFUPG 80/93, June 1993}
Topological excitations play a fundamental role in statistical mechanics and
field theory. They often drive phase transitions, as is the case in the
two-dimensional XY model \ref\ber{V. L. Berezinskii, Sov. Phys. JETP 32
(1971) 493, 34 (1972) 610; J. Kosterlitz and D. Thouless, J. Phys. C6 (1973)
1181; J. Villain, J. Physique 36 (1975) 581; R. Savit, Phys. Rev. B17 (1978)
1340.} and they can lead to drastical modifications of the perturbative
infrared behaviour of a theory \ref\pol{A. M. Polyakov, Phys. Lett. 59B
(1975) 82, Nucl. Phys. B120 (1977) 429, "Gauge Fields and Strings", Harwood
Academic Publishers, London (1987).}. The origin of these non-perturbative
phenomena lies in the compactness of a symmetry underlying a given model.
The topology of the gauge group is thus of paramount importance in Abelian
gauge theories, where the non-compact group $R$ and the compact group $U(1)$
lead to the same perturbation series. In three Euclidean dimensions,
for example, the compactness of $U(1)$ leads to the existence of instanton
solutions
of the Maxwell equations, which coincide with the familiar Dirac
magnetic monopoles \ref\god{For a review see: P. Goddard and D. Olive,
"Magnetic Monopoles in Gauge Field Theories", Rep. Prog. Phys. 41 (1978) 91.}
of three-dimensional Minkowski space. As was shown in \pol , in the weak
coupling limit these lead to the {\it confinement} of electric charges, by
effectively transforming the Coulomb potential from logarithmic to linear.

Three-dimensional space-times are characterized by the possibility of
adding a gauge invariant, non-conventional {\it Chern-Simons}
term to the gauge field action
\ref\des{R. Jackiw and S. Templeton, Phys. Rev. D23 (1981) 2291; J. Schonfeld,
Nucl. Phys. B185 (1981) 157; S. Deser, R. Jackiw and S. Templeton, Phys. Rev.
Lett. 48 (1982) 975, Ann. Phys. 140 (1982) 372.}. The resulting theory, with
Lagrangian density
\eqn\mcs{{\cal L}_{MCS} ={-1\over 4e^2}
F_{\mu \nu}F^{\mu \nu}+{\kappa \over 2}\
\epsilon^{\mu \nu \alpha }A_{\mu }\partial _{\nu }A_{\alpha }}
(in Minkowski space-time)
describes {\it massive photons} with mass $m=\kappa e^2$
(note that in $2+1$ dimensions $e^2$ has the dimension of mass). The Coulomb
interaction between charges minimally coupled to \mcs \ is correspondingly
exponentially screened. Therefore, the weak coupling non-perturbative behaviour
of the compact Maxwell-Chern-Simons theory depends crucially on
the interplay between magnetic monopoles and the Chern-Simons term,
since they clearly work in opposite
directions: the former promotes confinement of electric charges while the
latter screens the Coulomb interaction. The outcome of this competition can
have profound physical consequences, given that Maxwell-Chern-Simons theories
appear as effective theories for the fractional quantum Hall effect and for
two-dimensional spin models possibly relevant to
high-$T_c$ superconductivity \ref\fra{For a review see: E. Fradkin, "Field
Theories of Condensed Matter Systems", Addison-Wesley, Redwood City (1991).}.

There are two simple ways to analyze compact Abelian gauge theories. As was
pointed out in \pol , one automatically obtains the compact $U(1)$ group by
spontaneous breakdown of a compact non-Abelian group. Alternatively, one
can formulate the $U(1)$ model on a lattice, with the gauge fields being
{\it phases} of link variables.
Monopoles in compact Maxwell-Chern-Simons theory were
studied in \ref\pis{R. D. Pisarski, Phys. Rev. D34 (1986) 3851.}\
\ref\aff{I. Affleck, J. Harvey, L. Palla and G. Semenoff, Nucl. Phys. B328
(1989) 575.} by using the first of the above described approaches. It was
found that finite action single-monopole solutions do not exist. In \pis \
it was however shown that there exists a {\it complex} solution corresponding
to a monopole-antimonopole pair: this has real, finite
action proportional to the
distance between the monopole and the antimonopole. In \aff \
the absence of isolated monopoles was established by a perturbative
treatment of the Chern-Simons term.
All this indicates that monopoles are {\it linearly confined} by a string
of magnetic flux due to the Chern-Simons term.

It is therefore important to confirm and clarify this mechanism by analyzing
compact Maxwell-Chern-Simons theory on the lattice. Moreover, from this
approach one should also obtain the interaction between the strings.
While {\it non-compact}
versions of lattice Maxwell-Chern-Simons theory have been previously
studied \ref\fro{J. Fr\"ohlich and P. Marchetti, Comm. Math. Phys. 121 (1989)
177; E. Fradkin, Phys. Rev. Lett. 63 (1989) 322, Phys. Rev. B42 (1990) 570;
P. L\"uscher, Nucl. Phys. B326 (1989) 557; V. F. M\"uller, Z. Phys. C47 (1990)
301; D. Eliezer and G. Semenoff, Phys. Lett. B286 (1992) 118, Ann. Phys.
217 (1992) 66.},
no {\it compact} formulation of the same lattice models is yet available.
The difficulty lies in the explicit appearance of the gauge potential
$A_{\mu }$ in the Chern-Simons term. This makes it difficult to formulate
a periodic and gauge invariant lattice action with the correct continuum
limit. Here we avoid this problem by considering an equivalent formulation
of Maxwell-Chern-Simons theory in terms of {\it gauge invariant quantities}.

In \ref\jac{S. Deser and R. Jackiw, Phys. Lett. 139B (1984) 371.} it was
shown that the self-dual model
\eqn\sd{{\cal L}={1\over 2e^2} f_{\mu }f^{\mu }-{1\over 2me^2}
\epsilon ^{\mu \nu \alpha }f_{\mu }\partial _{\nu }f_{\alpha }\ ,}
considered first in ref.
\ref\tow{P. K. Townsend, K. Pilch and P. van Nieuwenhuizen, Phys. Lett.
136B (1984) 38.}, is {\it equivalent} (by a Legendre transformation) to the
Maxwell-Chern-Simons theory \mcs.
Indeed, the field equations obtained by varying \sd \  with respect to
 $f_\mu$ are identical to
 the Maxwell-Chern-Simons field equations if $f_\mu$ is
interpreted as the dual of the
field strength $F_{\mu \nu }$: $f^{\mu }=\epsilon^{\mu \alpha
\beta }F_{\alpha \beta }/2$.
This is possible since the equations of motion imply the ``Bianchi
identity'' $\partial_\mu f^\mu = 0$. Although this is valid only on shell,
this is sufficient to guarantee the equivalence of the {\it free} theories
\mcs \ and \sd, as shown in \jac.
We are thus led to consider the continuum model defined by the
Euclidean partition function
\eqn\pf{\eqalign{Z &= \int {\cal D}f_{\mu }
\ {\rm e}^{-S_E(f_{\mu })}\ ,\cr
S_E(f_{\mu }) &= \int d^3x\  {1\over 2e^2} \left( f_{\mu }f_{\mu }+{i\over m}
\epsilon_{\mu \nu \alpha }f_{\mu }\partial _{\nu }f_{\alpha }\right) \ .\cr}}

With this formulation at hand it is easy to construct a compact lattice
model of Maxwell-Chern-Simons theory. To this end we consider a cubic lattice
with lattice spacing $a$ and lattice sites denoted by the vectors ${\bf z}$.
The forward and backward lattice derivatives are defined as
\eqn\der{d_{\mu }g({\bf z}) \equiv {g({\bf z}+\hat {\bf \mu }a)-
g({\bf z})\over a} \ , \qquad
\hat d_{\mu }g({\bf z}) \equiv {g({\bf z})-g({\bf z}-\hat {\bf \mu }a)
\over a} \ ,}
where $\hat {\bf \mu }$ denotes a unit vector in the $\mu $ direction.
Correspondingly,
summation by parts on the lattice interchanges the two derivatives.

As usual \ref\kog{For a review see:
J. B. Kogut, Rev. Mod. Phys. 55 (1983) 775.}
we associate with each link $({\bf z}, \mu )$ of the lattice a real variable
denoted by $f_{\mu }({\bf z})$, which we identify (on shell)
as the dual field strength of a gauge
theory. If this underlying gauge theory is compact,
the link variables $f_{\mu }({\bf z})$
have to be considered as {\it angular variables} defined in the interval
$[-\pi/a^2, \pi/a^2]$, i.e. one has to identify
\eqn\com{f_{\mu }({\bf z}) \equiv f_{\mu }({\bf z}) +
{2\pi n_{\mu }({\bf z}) \over a^2}\ ,\qquad
n_{\mu }({\bf z}) \in Z\ .}
Using the topological mass m one can construct a second link variable with
dimension $[{\rm mass}^2]$,
${1\over m}\epsilon _{\mu \nu \alpha}d_{\nu }f_{\alpha }$,
the periodicity of which depends on the dimensionless parameter $am$.
In order to construct a model with a consistent continuum limit we choose a
lattice spacing satisfying
$1/ma=2n$,
with $n$ an integer.
As a consequence we have
\eqn\eps{{1\over 2m}\epsilon _{\mu \nu \alpha}d_{\nu }f_{\alpha }
({\bf z})\equiv {1\over 2m}\epsilon _{\mu \nu \alpha}d_{\nu }f_{\alpha }
({\bf z})+{2\pi h_{\mu }({\bf z})\over a^2}\ , }
where $h_\mu ({\bf z})$ are integer multiples of $n$.

In order to formulate our model we first note that the action
in \pf \ can be written as a sum of squares via the decomposition
\eqn\squ{S_E(f_{\mu }) =\int d^3x\ {1\over 2e^2} f_{\mu}^2+
{i\over 4e^2}\left( f_{\mu }+{1\over 2m}\epsilon_{\mu \nu \alpha}
\partial _{\nu }f_{\alpha }\right) ^2-{i\over 4e^2}\left( f_{\mu }
-{1\over 2m}\epsilon_{\mu \nu \alpha}\partial_{\nu }f_{\alpha}\right)^2 \ .}
We can now obtain a Villain-type model\ \ref\sav{For a review see: R. Savit,
Rev. Mod. Phys. 52 (1980) 453; H. Kleinert, "Gauge Fields in Condensed Matter
Physics", World Scientific, Singapore (1989).}
by summing over three sets of integer link variables,
whose purpose is to take into account the periodicity of
the link variables appearing in the three squares. Since we are integrating
over a unique fundamental variable $f_{\mu }$, however, it turns out that
two sets of integers are sufficient to enforce periodicity.
We therefore posit the following compact, lattice regularized version of the
Euclidean model \pf :
\eqn\mod{\eqalign{Z_L=\sum_{\{l_{\mu }\} \atop \{k_{\mu }\}} & \int_{-{\pi
\over a^2}}^{{\pi \over a^2}} {\cal D}f_{\mu }\  {\rm exp }
\sum_{{\bf z}, \mu } \left\{
-{a^3\over 2e^2}\left( f_{\mu }+{2\pi \over a^2}l_{\mu } \right) ^2 +
\right. \cr
&-{ia^3\over 4e^2}\left( f_{\mu }+{1\over 2m} \epsilon_{\mu \nu \alpha}
d_{\nu }f_{\alpha }+{2\pi \over a^2}(2l_{\mu }-k_{\mu })\right)^2 +\cr
& \left. +{ia^3\over 4e^2}\left( f_{\mu }-{1\over 2m} \epsilon_{\mu \nu \alpha}
d_{\nu }f_{\alpha }+{2\pi \over a^2}k_{\mu }\right)^2 \right\} \ ,\cr}}
where the sum runs over all lattice sites ${\bf z}$ and all directions $\mu $,
$l_{\mu }$ and $k_{\mu }$ are integer link variables
and we have introduced the notation ${\cal D}f_{\mu }=\prod _{{\bf z}, \mu }
df_{\mu }({\bf z})$. This partition function is clearly invariant under the
transformations \com , since these can be absorbed by a redefinition
$l^{\prime }_{\mu }\equiv l_{\mu }+n_{\mu }$ and
$k^{\prime }_{\mu }\equiv
k_{\mu }+n_{\mu }-{1\over 2am}\epsilon _{\mu \nu \alpha }ad_{\nu }n_{\alpha }$.
The formal continuum limit of the above lattice model is obtained by letting
$a\to 0$ and $\sum_{{\bf z}}\to (1/a^3)\int d^3{\bf z}$. As we now show,
in this limit we recover the continuum model \pf \ times a partition
function describing the topological excitations due to the compactness of
the underlying gauge symmetry.

Indeed, we now rewrite \mod \ in a fashion wich exposes explicitly these
{\it topological excitations} and their interactions. To this end we
decompose $k_{\mu }$ as
\eqn\dec{k_{\mu }\equiv l_{\mu }-{1\over 2ma}\epsilon_{\mu \nu \alpha}
ad_{\nu }l_{\alpha }+j_{\mu } \ ,}
with $j_{\mu }$ integers.
Correspondingly, the sum over all configurations $\{k_{\mu }\}$ in \mod \
can be replaced by a sum over all configurations $\{j_{\mu }\}$:
\eqn\cha{\eqalign{Z_L=\sum_{\{l_{\mu }\} \atop \{j_{\mu }\}}
 \int_{-\pi \over a^2}^{\pi \over a^2} {\cal D}f_{\mu } \ {\rm exp}
\sum _{{\bf z}, \mu } & \left\{ -{a^3\over 2e^2} g_{\mu }^2
-{ia^3\over 4e^2}\left( g_{\mu }+{1\over 2m}\epsilon _{\mu \nu \alpha}d_{\nu }
g_{\alpha }-{2\pi \over a^2}j_{\mu }\right) ^2+ \right. \cr
& \left. +{ia^3\over 4e^2}\left( g_{\mu }-{1\over 2m}
\epsilon_{\mu \nu \alpha}d_{\nu }
g_{\alpha }+{2\pi \over a^2}j_{\mu }\right) ^2 \right\} \ ,\cr}}
where $g_{\mu }\equiv f_{\mu }+(2\pi /a^2)l_{\mu }$. By changing variables
from $f_{\mu }$ to $g_{\mu }$ in the integration and performing the sum over
all configurations $\{l_{\mu }\}$ we obtain
\eqn\gau{Z_L=\sum_{\{j_{\mu }\}} \int_{-\infty }^{+\infty } {\cal D}
f_{\mu } \ {\rm exp} \sum_{{\bf z}, \mu }
\left\{ {-a^3\over 2e^2}f_{\mu }\left(
\delta _{\mu \alpha }+{i\over m}\epsilon_{\mu \nu \alpha}d_{\nu } \right)
f_{\alpha } +{i2\pi a\over e^2}f_{\mu }j_{\mu } \right\} \ .}

In a last step we perform the Gaussian integration over $f_{\mu }$. To this
end we note that, by a summation by parts, the operator appearing in the
quadratic term in $f_{\mu }$ can be rewritten as
\eqn\nop{\delta _{\mu \alpha}+{i\over m} \epsilon_{\mu \nu \alpha}
 d_{\nu } = \delta _{\mu \alpha}+{i\over m}\epsilon_{\mu \nu \alpha}
D_{\nu }\ , \quad
D_{\mu } \equiv {1\over 2}\left( d_{\mu }+\hat d_{\mu }\right) \ .}
Its inverse is given by
\eqn\inw{\eqalign{G_{\mu \nu}({\bf z}, {\bf z}^{\prime }) & = \left( m^2
\delta _{\mu \nu }-D_{\mu }D_{\nu }-im\epsilon_{\mu \alpha \nu}D_{\alpha }
\right) \ G({\bf z}-{\bf z}^{\prime }) \ ,\cr
& \left( -D_{\mu }D_{\mu }+m^2 \right) G({\bf z}-{\bf z}^{\prime })
= \delta _{{\bf z}, {\bf z}^{\prime }} \ .\cr }}
Note that the operator $D_{\mu }D_{\mu }$ appearing in this formula
represents a lattice regularized version of the Laplacian. Using \inw \
we obtain our final result:
\eqn\fin{Z_L=Z_0 \sum_{\{j_{\mu }\}} {\rm exp} \sum _{{{\bf z}, \mu } \atop
{{\bf z}^{\prime }, \nu }} \ -{4\pi ^2\over 2ae^2} \ j_{\mu }({\bf z})
G_{\mu \nu }({\bf z}-{\bf z}^{\prime }) j_{\nu }({\bf z}^{\prime }) \ ,}
where
\eqn\fre{Z_0=\int {\cal D}f_{\mu } \ {\rm exp} \sum _{{\bf z}, \mu }
\left\{ -{a^3\over 2e^2}f_{\mu }f_{\mu }-{ia^3\over 2me^2}f_{\mu }
\epsilon_{\mu \nu \alpha}d_{\nu }f_{\alpha } \right\} \ ,}
is the Gaussian partition function for a free massive photon on the lattice.
As anticipated, this reduces to the continuum partition function \pf \ in
the formal continuum limit.
The remaining factor in \fin \ describes the topological excitations, which
we now discuss.

First, let us note that these topological excitations are strings: these can
be closed (rings), in which case $a\hat d_{\mu }j_{\mu }=0$, or open, in which
case we identify the integers
\eqn\ide{a\hat d_{\mu }j_{\mu }=q}
with the {\it magnetic monopoles}. In order to justify this interpretation
let us consider the continuum limit of the topological excitations.
The continuum field equations,
 derived from the continuum action \pf , are given by
\eqn\fic{f_{\mu }+{i\over m}\epsilon_{\mu \nu \alpha }\partial _{\nu }
f_{\alpha }=0 \ .}
When $f_\mu$ is interpreted as
 the dual field strength of a
{\it compact} gauge theory, however, it can contain {\it string
singularities} \pol \ $j_\mu$: $f_{\mu }  =f_{\mu }^{\rm reg} - 2\pi j_\mu$
, where $j_\mu$ \ are of the type $j_{\mu }=n \ (\theta(x_0-L/2)-
\theta(x_0+L/2))\delta(x_1)\delta(x_2)$, with $n$ an integer.
These string singularities can be brought to the right-hand side of \fic ,
where they act as sources for the regular part of $f_{\mu }$,
\eqn\ceq{f_{\mu }^{\rm reg} +{i\over m}
\epsilon_{\mu \nu \alpha}\partial _{\nu }
f_{\alpha }^{\rm reg} =2\pi j_{\mu } \ .}
Inserting \ceq \ and its inverse
\eqn\inv{f_{\mu }^{\rm reg} =2\pi {m^2\delta _{\mu \nu}
-\partial _{\mu }\partial _{\nu }
-im\epsilon_{\mu \alpha \nu}\partial _{\alpha } \over -\partial _{\mu }
\partial _{\mu }+m^2}\ j_{\nu }\ ,}
in the action $S_E(f_{\mu })$ in \pf \ we obtain
\eqn\cmo{S_{{\rm Top}}=\int d^3x\ {4\pi ^2\over 2e^2}\ j_{\mu }
{m^2\delta _{\mu \nu}-\partial _{\mu }\partial _{\nu }-im
\epsilon_{\mu \alpha \nu}\partial _{\alpha }\over -\partial _{\mu }
\partial _{\mu }+m^2}j_{\nu }\ ,}
which is exactly the continuum limit of the topological action in \fin .

Eq. \inv \ represents the {\it Maxwell-Chern-Simons monopole solution}.
The fact that $\partial _{\mu }f_{\mu }^{\rm reg} =2\pi \partial _{\mu }j_{\mu
}
=2\pi \ {\rm integer} \ \delta ^3(x-x_{mon})$
justifies our identification \ide \ in the lattice model.
As expected, the magnetic field is exponentially screened with
a characteristic length $(1/m)$ determined by the topological
Chern-Simons mass.
In the limit $m \to 0$, \inv \ reduces to the familiar monopole  solution
of compact, (2+1)-dimensional QED \pol .
Correspondingly \cmo \ reduces to the action for a Coulomb gas of monopoles
\pol .
Our monopole solution \inv \ is {\it different} from previously considered
Abelian solutions \ref\hen{M. Henneux and C.
Teitelboim, Phys. Rev. Lett. 56 (1986) 689.}\ \pis\ which describe long
range magnetic fields.
In analogy to Pisarski's non-Abelian
solution \pis , our Abelian solution is complex; however the
corresponding action contains a positive definite real part.

Let us now return to our lattice model and
consider the action for a monopole-antimonopole pair united by a
string of length $L$. This contains a positive piece given by
\eqn\mam{S_{M\bar M}= {4\pi ^2m^2\over 2e^2a^2} G(0)\ L \ ,}
with $G$ defined in eq. \inw .
This shows that a single-monopole solution $(L\to \infty )$ has infinite
action; therefore isolated monopoles are completely suppressed in the
partition function. Monopoles can only appear as monopole-antimonopole pairs
{\it linearly confined} by a string; since $2e^2$ plays the role of
temperature and $G(0)=a^2 g(ma)$
we can identify $m^2 4\pi ^2g(ma)$ as the corresponding
{\it string tension}.

The string singularities \ceq \ ,
due to the compactness of the model, correspond exactly to
the {\it induced current} first postulated in \hen. The necessity of such an
induced current in the presence of a monopole was also recognized by Pisarski
\pis , who first advanced an explanation  of its physical meaning.
As is by now well know, in topologically massive quantum electrodynamics,
external currents generate magnetic flux due to the presence of $f_{\mu }$
in the equations of motion. What happens in presence of a monopole is the
converse: magnetic flux generates an induced current which flows through
the Dirac string. Due to its electric charge, this becomes observable and
acquires a finite energy per unit length. Due to this energy, configurations
with infinitely long strings are suppressed.
This mechanism is clearly exposed in eqs. \fin \ and \cmo\ .
The interaction energy for the strings contains three terms: the first is
the electric-electric interaction between the charged strings, the second is
the magnetic-magnetic interaction between the monopoles at the end of the
strings, the third is the electric-magnetic interaction between the
electric charge of one string and the magnetic field induced by the second
via the Biot-Savart law.
All this is reminescent of self-dual $Z_N$ models in even dimensions
 \ref\car{J. L. Cardy and E. Rabinovici, Nucl. Phys. B205 (1982) 1.},
 the difference being that here it is the same objects that
carry both the electric and the magnetic charge.
Note also, that the parity-violating, imaginary electric-magnetic
interaction contains a topological term measuring the {\it (self-) linking
number} of closed strings  \ref\poly{A. M. Polyakov, Mod. Phys. Lett. A3 (1988)
325.}.

A word of care is due at this point. Indeed, the fact that $\partial_{\mu }
j_{\mu }\ne 0$ does not imply any violation of charge conservation.
In fact $f_{\mu }^{\rm reg }$ appearing on the left-hand side
of \ceq \ can not be interpreted as the dual field strength of a gauge theory
exactly because it is stripped off its singularities. Correspondingly,
$j_{\mu }$ is not a current minimally coupled to a gauge theory and need
therefore not be conserved. As is shown in \hen \ \pis , the total current
minimally coupled to $A_{\mu }$ in the gauge formulation is indeed conserved.

Let us now write the field equations of topologically massive electrodynamics
in the form
\eqn\can{{1\over e^2}\epsilon_{\mu \nu \alpha}\partial _{\nu }f_{\alpha }
-{im\over e^2}f_{\mu }=J_{\mu }^P \ .}
The current appearing in this formulation is the particle-number current
density. Comparing this with \ceq \ we recognize first that $J_{\mu }^P$
is imaginary in Euclidean space. In Minkowski space it would be real and
$\partial _{\mu }J_{\mu }^P\ne 0$ would correspond to the creation of
particles at the event of the monopole \hen . Comparing with the divergence of
\ceq \ we recognize that the number of these  particles is quantized
only if the Chern-Simons coupling constant is quantized as
$2\pi \kappa \equiv 2\pi m/ e^2 = {\rm integer}$,
in full agreement with \hen \ \pis .

We conclude by remarking that the topological excitations we have found
in compact Maxwell-Chern-Simons theory are the same as the corresponding
ones in the compact Abelian Higgs model in (2+1) dimensions \ref\ein{ M. B.
Einhorn and R. Savit, Phys. Rev. D17 (1978) 2583, Phys. Rev. D19 (1979) 1198.}.
There, the same mechanism takes place: coupling the gauge fields to a scalar
effectively attaches flux tubes to the monopoles, which become linearly
confined. Indeed, the real part of the interaction between strings is
identical in the two models; the only difference lies in the
parity-violating imaginary
part of the interaction, present in the Maxwell-Chern-Simons model.
We postpone the analysis of the phase structure of our model to
a forthcoming publication.

\bigbreak\bigskip\bigskip\centerline{{\bf Acknowledgements}}\nobreak
We greatly profited from discussions with L. Alvarez-Gaum\'e. M. C. D.
thanks the theory division of CERN, where this work was initiated, for
hospitality. This work was partially supported by M.U.R.S.T..

\listrefs
\end